\documentclass[11pt]{article}
 \usepackage{graphicx,amsmath,eucal}
\def\bz{\mathbf{Z}}
\def\an{\mathbf{A}^n}
\def\ann#1{\mathbf{A}_#1}
\def\f{\mathbf{F}}
\def\cm{\mathcal{M}}
    \def\xdownarrow#1{%
    {\left\downarrow\vbox to #1{}\right.\kern-\nulldelimiterspace}
      }
\def\jonq{Jonqui\`eres\ }
\long\def\br#1#2@{\par\vskip6pt\noindent\textbf{Remark #1}\quad#2}
\begin{document}
\begin{center}
    {\large\bf Symmetric Key Encryption for Arbitrary Block Sizes\\[6pt]
    from  Affine    Spaces}
        \\[6pt]
    \textbf{P Vanchinathan}\\[2pt]
VIT University\\
Vandalur--Kelambakkam Road \\
Chennai 600127, INDIA\\ \tt  vanchinathan.p@vit.ac.in
\end{center}

\noindent{\textbf{Abstract:}\quad} A symmetric key encryption scheme is
described for blocks of general size $N$ that is a product of powers of many
prime numbers. This is accomplished 
by realising each number  (representing a message
unit) as a point in a product of affine spaces over various finite fields. Then
algebro-geometric transformations on those spaces is transported back to
provide encryption. For a specific block size${}<2^{128}$ we get  more
than $2^{5478}$ keys.

\vspace{6pt}
\section*{Introduction}

A typical block cipher  
takes the set $\cm=\{0,1,2,\ldots, N-1\}$, the set of message units, 
and provides one permutation
of that set corresponding to each element (\textit{a secret key}) of another set
called the keyspace. 
The larger the keyspace,  the more difficult it is to break the cryptosystem by exhaustive search for keys.

For ease of implementation in electronic hardware, usually  the cardinality $N$
is taken as  a power of 2. When $N=2^n$, these are called $n$-bit block ciphers.

The well-known  industry standard cipher,  
AES encryption,  provides three variants that are 
128-bit, 192-bit and 256-bit block ciphers. (That is, they permute sets of huge
sizes such as $2^{128}$).

There are situations where one needs permutations of sets of size different 
from powers of 2, and that branch of cryptology is called
\textit{format-preserving encryption} (FPE).  (See Black and Rogaway [BR02],  Brightwell and Smith [BS97] and 
[FIPS74]). FPE, for example, tries to encrypt a 16-digit credit card number
into  something that  again looks like a credit card number.

In this paper we describe  a construction of  block cipher for sets of size $N$
with $N$  factorizable into  many prime powers.

The novelty in our scheme is the 
choice of geometric model for the set of message units. It is a product
of \textit{affine spaces}. Being an uncomplicated affine variety
the task of mounting points of the message space onto  this geometric object
is a simple one.
The transformations on our geometric model arise from \jonq automorphisms: they
are our preferred choice due to their ready invertibility though there are
different kinds of automorphisms there.

We simultaneously use \jonq automorphisms over 
all finite fields $\mathbf{F}_p$  corresponding to every prime divisor $p$ of the block size $N$, 
and then patch them to yield a permutation on the block by appealing to Chinese Remainder Theorem.

This encryption scheme makes available an immense collection of permutations 
(i.e., a huge keyspace). To
compare, while AES  provides $2^{128}$ keys, this  method 
for a  slightly smaller block size (but chosen conveniently), 
even with an artificial restriction  to 
smaller keys (\textit{``polynomials of degree${}\leq5 $''}), yields a keyspace
of size  bigger  than $2^{5478}$. Moreover,  these permutation calculations can
be parallelized for speed.

Here are the salient features of this encryption scheme:
\begin{itemize}
\item 
        As we map the message units  with the 
        points of affine  spaces,
        one layer of complexity  at pre-encryption stage is eliminated
        providing considerable simplification.
    \item 
    Affine spaces as opposed to elliptic curves and  abelian varieties are not only conceptually simpler objects of algebraic
    geometry but also 
        come with a vast collection of automorphisms compared to them,
        thereby affording a large  keyspace  meeting an important
        requirement of information security.
\item Block ciphers for blocks of size different from powers of a single prime are provided.
\item All the bijections and permutations used here  are explicit,
      natural mathematical constructions, and they are 
        easily implementable as computer programs. (We have 
        implemented this algorithm for  numbers of the form
        $N=p^4q^4$ as a Python program with 150 lines of code).
\item Computations are parallelizable: for any permutation provided by this
    scheme, where any two different elements should be sent can be computed without the knowledge of each other. 
\end{itemize}
Now we hasten to add that though theoretical description is available for
blocks whose  size $N$ has two or more  distinct prime divisors, from the 
viewpoint of utility value, only 
a certain restricted numbers $N$ might be suitable. We provide such a suitable number 
to compare the keyspace size  with 128-bit encryption.

For convenience and for avoiding notational clutter  we  describe our scheme in
the simpler case  where  block size $N$ is of the form $N=p^rq^s$, with $p,q$
distinct primes. Working out the general case  of $N$   having three 
or more prime factors is straightforward.

This paper is organized as follows: 
\begin{itemize}
    \item
In Section 1  we recall the definition of \jonq
automorphisms which is a  fundamental ingredient in our scheme. This focusses
on sets whose cardinality is a prime power. 
        \item In the second section
we provide the full description of our encryption scheme which builds on top of
Section 1 and uses Chinese Remainder Theorem for weaving together 
        the automorphisms
coming from various prime-power divisors of the block size.
\item
In Section 3 an alternative description is provided in an algorithmic fashion
to aid in computation.
 \item
 Finally in Section 4 we illustrate with an example that uses
 blocks of a specific size slightly less than $2^{128}$ and  compute the size
        of the key space. 
\end{itemize}
\section{Jonqui\`eres Automorphisms}

Consider the affine space $\an$ of dimension $n$,
the set  of $n$-tuples of elements
over some field $\f$. 
 We avoid calling it a vector space for the simple reason that we will be doing non-linear operations on this set.

We use $(x_1,x_2,\ldots, x_n)$ as co-ordinates of a point in this space.
As an affine algebraic variety $\an$ admits many automorphisms, besides the
translations and linear automorphisms.  \jonq  has defined
a `triangular' family of polynomial automorphisms ([J1864]).  To define a \jonq automorphism one has to first choose $n-1$ polynomials
 $P_1, P_2,\ldots, P_{n-1}$ with coefficients in $\f$ such that  $P_1$ involves  just one variable $x_1$, $P_2$  involves  just $x_1$  and $x_2$. In general the polynomial  $P_i$ is taken to be involving only the first $i$ variables $x_1,x_2,\ldots, x_i$.
Along with this we also need non-zero scalars from the base field, 
$a_i\in \mathbf{F}^*$, for $1\le 1\le n$.

\noindent\textbf{Lemma 1} (\textbf{E. de Jonqui\`eres, 1864}) \textit{ 
Assume $a_i$ and the polynomials $P_i$ are as above. Further assume that
for any of these $P_i$ the degree is less than $p$. Then the
map $J\colon \an\to\an$  defined by  
sending $(x_1,x_2,,\ldots, x_n)$ to $(y_1, y_2,\ldots, y_n )$
by the formula 
\[\begin{array}{rcl}y_1&=&a_1x_1\\  y_2 &=&a_2x_2 +P_1(x_1)\\
    y_3&=&a_3x_3+P_2(x_1,x_2)\\ \vdots &\vdots&\qquad\vdots\\
y_n&=&a_nx_n+P_{n-1}(x_1,x_2,\ldots, x_{n-1})\end{array}\]
is an automorphism of the affine space as an affine algebraic variety. }

\vspace{6pt}
\textit{Proof:\quad}
\textit{The condition on the degree is to avoid  
terms of the form $x_i^p$ which bring inseparability issues. In all real world cryptographic applications
we need to choose $p>100$, and degrees $< 10$}.

To show that this is an automorphism one should exhibit an inverse function and show that the inverse is also a polynomial function.
Inverting is not difficult, one simply imitates the procedure for  solving a triangular system of \textit{linear} equations. 

The inverse $J^{-1}$ can be computed \textit{sequentially} by the formulas
below:
\[J^{-1}(y_1,y_2,\ldots,y_n) = (x_1,x_2,\ldots,x_n)\quad\mbox{with }\]
\[\begin{array}{rcl}
x_1&=&a_1^{-1}y_1\\[2pt]  x_2&= &a_2^{-1}\big(y_2-P_1(x_1)\big)\\ [2pt ]  x_3 &=& a_3^{-1}\big( y_3-P_2(x_1,x_2)\big)\\
&\vdots\\  x_n&=& a_n^{-1}\big(y_n-P_{n-1}(x_1,x_2,\ldots,x_{n-1})\big)\end{array}\] 
This proves the lemma and much more: the inverse of $J$ is also
a \jonq automorphism (see the first remark below).
\br1
The data consisting of scalars $a_i$'s and polynomials $P_i$'s essentially 
form the encryption key for our scheme.  To justify the name \textit{symmetric key} encryption
we can see that in the opposite direction the scalars $b_i$'s are simply
the inverses of $a_i$'s modulo $p$, and the polynomials in the opposite
directions are got by changing the signs  of all the coefficients of $P_i$'s
followed by multiplication by $b_i$'s. The upshot of this is that the decryption
key is readily obtained from the encryption key and they are mutual inverses as
required.@
\br2 
Other than the fact that these $P_i$'s should involve only the
variables $x_1$ to $x_{i}$ (\textit{backward-mixing}) there is no restriction
on them in order to define an automorphism. So one has a huge collection of
\jonq automorphisms that can be readily written down. (see Lemma
3 below).@
\br3 For practical considerations to enable good mixing
one should also introduce \textit{forward-mixing}. For this purpose we take two
such \jonq automorphisms, apply  one of them first,
follow it up with  the  reversal map, $(x_1,x_2,\ldots,x_{n-1},x_n)\mapsto
(x_n,x_{n-1},\ldots,x_2,x_1 )$ and then apply the second \jonq automorphism.
(Alternatively this can be understood as analogous to carrying out an upper 
triangular transformation followed by a lower triangular transformation).@
\br4 Any function $\f_p\to \f_p$ is a polynomial function by Lagrange
interpolation formula. Higher dimensional analogue of this is also true. 
Any function from $\mathbf{A}_p^n$ to itself is a polynomial, actually a
polynomial of degree at most $(p-1)^n$.  I thank \texttt{user9072}  
of the internet forum \texttt{www.mathoverflow.com} for pointing out the validity of this
generalization.@

\vspace{1pc}
\noindent\textbf{Lemma 2}\quad Let 
$\phi_1=(F_1,F_2,\ldots,F_n)\colon \ann p^n\to \ann p^n$ and
$\phi_2=(G_1,G_2,\ldots,G_n)\colon\allowbreak
\ann p^n\to \ann p^n$ 
be two (polynomial) automorphisms.
If total
degree${} <p$ for every $F_i$ and $G_j$ and if $\phi_1,\phi_2 $ are
different as polynomials (i.e., at least some pair of corresponding 
coefficients are different) then $\phi_1\neq\phi_2$ as \textit{functions}
on $\ann p^n $.

\vspace{6pt}
\textit{Proof:\quad} Suppose $\phi_1$ and $\phi_2$ are one and the same as
functions.
Then, in particular, $F_1(x_1,x_2,\ldots,x_n)\equiv G_1(x_1,x_2,\ldots,x_n)$.
Specializing all the $x_j,\  j\geq2$ at 1, we get the following equality 
of univariate polynomial functions over $\f_p$: $F_1(x_1,1,1,\ldots,1)=G_1(x_1,1,1,\ldots,1)$.
This shows that the difference between these polynomials in $x_1$ is of degree
at least $p$,  as it has all the elements of $\f_p$ as its roots. \hfill QED

\vspace{1pc}
\noindent\textbf{Lemma 3}\quad The number of  `upper triangular' 
\jonq automorphisms over a given prime
field using low degree polynomials are as in the table below:
\[
    \begin{array}{|l|l|l|} \hline 
\mbox{Affine Space} & \mathrm{Degree} & \mbox{Number of Automorphisms}\\ \hline
\ann p^4 & \leq 3 & p^{34}(p-1)^4 \vrule width0pt height13pt \\[2pt]
        \ann p^4 & \leq 4 & p^{55}(p-1)^4 \\[2pt]
        \ann p^4 & \leq 5 & p^{83}(p-1)^5 \\[2pt]
        \ann p^5 & \leq 3 & p^{69}(p-1)^5 \\[2pt]
        \ann p^5 & \leq 4 & p^{125}(p-1)^5 \\[2pt]
        \ann p^5 & \leq 5 & p^{209}(p-1)^5 \\
        \hline
    \end{array}
    \]

\textit{Proof}:\quad  
    This follows from the  well-known formula for the number of monomials
    of degree $d$  in $n$ variables.

\section{Construction of the Block Cipher}
As  stated in the introduction we take for simplicity  $N= p^rq^s  $.
Our object is  to produce explicitly a large family of  constructible
permutations of the set of message units, $\cm=\{0,1,2,\ldots, N-1\}$.

Our idea can be summarised as below: 
\begin{description}
 \item[Step (i)] to identify $\cm$  with  a geometric object, viz.\ 
a cartesian product of  affine spaces (over  different prime fields) through explicit bijections.
\item[Step (ii)] apply  \jonq automorphisms independently in each of the affine
    spaces forming the terms of  the cartesian product above.
\item[Step (iii)] 
 transport  the  product of \jonq automorphisms back to the message
        space $\cm$ through the inverse of the bijections mentioned in Step (i).
        It is simply retracing the bijections of Step (i).
\end{description}

\subsection{Details of Step (i): Message Space to Affine Space }
 For this  we regard $\cm$ as the commutative ring of integers modulo $N$. Our  
 bijection of Step (i)  is obtained as a composition of two bijections:
first one  denoted by $\psi$, is a ring isomorphism,  and the second one 
denoted by $\delta$ is a set-theoretic bijection
 as indicated below:
\[\cm=\bz/N\bz\stackrel{\psi}{\longrightarrow} \bz/p^r\bz\times \bz/q^s\bz\stackrel{\delta}{\longrightarrow }\mathbf{A}_p^r\times \mathbf{A}_q^s \]

\noindent \textbf{Description of }$\psi, \psi^{-1}$:\quad $\psi$ is simply the calculation of the two remainders of a number for division by $p^r$ and  $q^s$.
\[\psi(m) = \big( m\,\mathrm{mod}\,{p^r},\  m\,\mathrm{mod\,}{q^s}\big)\]
 The inverse, $\psi^{-1}$, is the map provided by Chinese Remainder Theorem.

\vspace{1pc}
\noindent \textbf{Description of }$\delta$:\quad 
First  we define  a function $\delta_p$ for  $a<p^r$ as a vector
formed by its digits in base $p$ expansion:
\[ \delta_p(a)=(\alpha_0,\alpha_1,\ldots, \alpha_{r-1}) \qquad\mbox{for }
a=\sum_{j=0}^r \alpha_j p^j,\
0\le \alpha_j<p.\]Similarly  $\delta_q(b)$ is defined  using base $q$
digits.
Now $\delta$ is defined as 
\[\begin{array}{ccc} \delta\colon \bz/p^r\bz\times \bz/q^s\bz
    &\longrightarrow & \mathbf{A}_p^r\times
    \mathbf{A}_q^s,\\[4pt]
\delta (a,b) &=& \big(\delta_p(a), \delta_q(b)\big)\end{array} \]

Inverse $\delta_p^{-1}$ is even easier to compute. Interpreting a  vector
with all components integers less than $p$ as the base-p digits, this 
will represent a number
less than $p^r$. (Similarly for $\delta_q^{-1}$).

\vspace{1pc}
\noindent \textbf{Example} 

\vspace{3pt}
\noindent Take $N=5000=2^35^4$. Let us first calculate $\psi(471)$. As
$471\pmod {2^3}=7$  and $ 471\pmod{5^4}=96$ we have
$\psi(471)=(7,96)$.

Now $\delta\big( (7,96)\big)= \big( (1,1,1), (0,3,4,1)\big)$
(because 7 in binary is 111 and 96 in base 5 is 341. As $5^4|N$, we have to
write residues modulo  $5^4$ as 4-digit numbers in base 5, inserting leading zeros where needed).

\subsection{Details of Step (ii): Geometric Transformation within the Affine
Spaces}
We  take 
two pairs of \jonq automorphisms $(J_p^1, J_q^1)$ and
$(J_p^2,J_q^2) $ which makes up our encryption key in this case.
Each pair  gives rise to a bijection $J=J_p\times J_q\colon
\mathbf{A}_p^r\times \mathbf{A}_q^s\to   \mathbf{A}_p^r\times \mathbf{A}_q^s,$
i.e.,
a cartesian product of \jonq automorphisms  $J_p\colon
\mathbf{A}^r_p\to \mathbf{A}^r_p$ , and $J_q\colon
\mathbf{A}^s_q\to\mathbf{A}^s_q$.    

\vspace{6pt}
\noindent\textbf{Key Selection}\quad To get a key one has to select   finite sequences (of
appropriate length) of random integers in the ranges $[0,p-1]$ and $[0,q-1]$
respectively to be used as coefficients of polynomials which make up
the four \jonq automorphisms, $ J_p^1,J_p^2,J_q^1,J_q^2$.

Let us assume such a selection  has been made giving rise to a key $K$ 
consisting of two pairs of \jonq automorphisms $K: (J_p^1,J_q^1),
(J_p^2,J_q^2)$. 
Define reversal map as 
$\mathbf{A}_p^r\times \mathbf{A}_q^s\to
\mathbf{A}_p^r\times \mathbf{A}_q^s$ by
\[ \big(\alpha_0,\alpha_1,\ldots,\alpha_{r-1};\
\beta_0,\beta_1,\ldots,\beta_{s-1}\big)\stackrel{\mathrm{rev}}{\longrightarrow}
    \big(\alpha_{r-1},\alpha_{r-2},\ldots,\alpha_0;\
\beta_{s-1},\beta_{s-2},\ldots,\beta_0\big) 
\]
We apply the first pair $(J_p^1, J_q^2)$ to an element of the product
$\mathbf{A}_p^r\times \mathbf{A}_q^s$. Then apply 
 reversal and finally  apply the second pair of \jonq automorphisms
$(J_p^2,J_q^2)$. This completes Step (ii).
 \subsection{Details of Step (iii): Back  to Message Space from Affine Space}
In this stage we travel backwards to message space.
\[
\mathbf{A}_p^r\times \mathbf{A}_q^s
\longrightarrow \bz/p^r\bz\times \bz/q^s\bz\longrightarrow
    \bz/N\bz=\cm
\]
 Given an element in $\ann p^r $ take the components as base $p$  digits and
 compute the number${}<p^r$ represented by it. Similarly do the same for
 the element  from $\ann q^s$ getting a number less than $q^s$.

 The second leg of our return journey is $\delta^{-1}$. By Chinese remainder
 theorem we get a unique number less than $N=p^rq^s$
 using the two numbers just obtained in the first leg of the journey taking us
 back  to $\cm$.
\subsection{The Full Picture}
Now we get an encryption, i.e., a permutation $E_K$ of  $\cm=\bz/N\bz$
corresponding to the choice of key  $K$,  through  a  sequence of compositions
as described by the commutative diagram shown in Figure 1. 

\begin{figure}[h]
\[
    \begin{array}{ccc}
    \bz/N\bz & \stackrel{\textstyle\psi}{\longrightarrow} & \bz/p^r\bz\times \bz/q^s\bz\\
        & & \bigg\downarrow\mbox{ \rlap{$ \delta$}} \\
    |    & & \ann p^r\times \ann q^s \\
        |    & & \bigg\downarrow \mbox{ \rlap{$ J_p^1\times J_q^1$}}\\
    |  \mbox{\rlap{ $E_K$}}    & & \ann  p^r\times \ann q^s \\
    |    & & \bigg\downarrow\mbox{ \rlap{rev}}\\
    |    & & \ann p^r\times \ann q^s \\
        |    & & \bigg\downarrow\mbox{\rlap{\ $J_p^2\times J_q^2$}}\\
  \bigg  \downarrow & & \ann p^r\times \ann q^s \\
         & & \bigg\downarrow\mbox{\rlap{\ $ \delta^{-1}$}} \\
    \bz/N\bz & \stackrel{\textstyle \psi^{-1}}{\longleftarrow} & \bz/p^r\bz\times \bz/q^s\bz\\
    \end{array}
 \]
 \caption{Schematic Description of the Encryption Algorithm}
\end{figure}

\section{Algorithm}
\textbf{Set-up:}\quad   $\cm=\{0,1,2,\ldots, N-1\}$,  with $N= p^rq^s$

\vspace{6pt}
\textbf{Encryption Key}:\quad Two ordered sets of polynomials
$\{P_1,P_2,\ldots, P_{r-1}\}$, and $ \{ Q_1,Q_2,\ldots, Q_{s-1}\}$, with the
first set having coefficients in $\f_p$, the second set in $\f_q$, with the
$i$th polynomial in both sets involving only  the first $i$ variables; and
scalars $a_i\in\f_p^*, b_j\in \f_q^*\ $ for $0\le i\le r-1,\ 0\le j\le s-1$.
Denote this half of  key datum by  $K$, i,e,,
\[K=(P_1,P_2,\ldots, P_{r-1},a_0,a_1,\ldots,a_{r-1},Q_1,Q_2,\ldots, Q_{s-1},b_0,b_1,\ldots,b_{s-1}).\]
Similarly the other half of key datum is:
\[K'=(P_1',P_2',\ldots, P_{r-1}',a_0',a_1',\ldots,a_{r-1}',Q_1',Q_2,'\ldots,
Q_{s-1}',b_0',b_1',\ldots,b_{s-1}').\]

\noindent \textbf{Keyspace}\quad 
 A single key for our encryption scheme,
  in  the general case where $N=p_1^{r_1}p_2^{r_2}\cdots p_k^{r_k}$ 
  comprises   two pairs of 
$k$-tuples of \jonq:
$(J_{p_1}^1,J_{p_2}^1,\ldots, J_{p_k}^1;\ 
J_{p_1}^2,J_{p_2}^2,\ldots,J_{p_k}^2)$ 
and 
$(F_{p_1}^1,F_{p_2}^1,\ldots, F_{p_k}^1;\allowbreak  
F_{p_1}^2,F_{p_2}^2,\ldots,F_{p_k}^2)$. 

\vspace{1pc}
\noindent 
\textbf{Algorithm}:  
Given a number $m\in \cm $  its encrypted value $E_K(m)$ for  given key $K$ is computed by the  steps below.
\begin{itemize}
\item  Calculate the remainders  $m_p, m_q$: \[m_p\equiv m\pmod {p^r};\qquad m_q\equiv m\pmod{q^s}\]
\item Extract the digits $\alpha_i$  of $m_p$ and $\beta_j$ of  $m_q$ in the bases $p$ and $q$ respectively. 
\[\sum_{i=0}^{r-1}\alpha_i p^i = m_p;\qquad  \sum_{j=0}^{s-1}\beta_jq^j = m_q;\qquad\qquad (\alpha_i< p,\ \ \beta_j< q \ \forall i, j)\]
\item Transform the alphas and betas  by respective \jonq automorphisms.
\[(\alpha'_0,\alpha'_1,\ldots,\alpha'_{r-1})= \big (a_0\alpha_0,\  a_1\alpha_1+P_1(\alpha_0),\   a_2\alpha_2 +P_2(\alpha_0,\alpha_1),\]
\[ {}\hskip9em\ldots, a_{r-1}\alpha_{r-1} + P_{r-1}(\alpha_0,\alpha_1,\ldots, \alpha_{r-2})\big)\]
\[(\beta'_0,\beta'_1,\ldots,\beta'_{s-1})= \big (b_0\beta_0,\  b_1\beta_1+Q_1(\beta_0),\   b_2\beta_2 +Q_2(\beta_0,\beta_1),\]
\[ {}\hskip9em\ldots, b_{s-1}\beta_{s-1} + Q_{s-1}(\beta_0,\beta_1,\ldots, \beta_{s-2})\big)\]
\item Reverse the components and apply the other pair of  \jonq
    automorphisms.
        \[(\alpha''_0,\alpha''_1,\ldots,\alpha''_{r-1})= 
        \big (a_0'\alpha_{r-1}',\
        a_1'\alpha_{r-2}'+P_1'(\alpha_{r-1}'),\   a_2'\alpha_{r-3}'
        +P_2'(\alpha_{r-1}',\alpha_{r-2}'),\] 
        \[ {}\hskip9em\ldots, a_{r-1}'\alpha_0' +
        P_{r-1}'(\alpha_{r-1},\alpha_{r-2},\ldots, \alpha_1)\big)\]
\[(\beta''_0,\beta''_1,\ldots,\beta''_{s-1})= 
        \big (b_0'\beta _{s-1}',\
        b_1'\beta _{s-2}'+P_1'(\beta _{s-1}'),\   b_2'\beta_{r-3}'
        +P_2'(\beta _{s-1}',\beta _{s-2}'),\] 
        \[ {}\hskip9em\ldots, b_{s-1}'\beta _0' +
        P_{s-1}'(\beta _{s-1},\beta _{s-2},\ldots, \beta _1)\big)\]

\item Assemble the individual digits $\alpha''_i$'s and $\beta''_j$'s into
    numbers $m_p'(<p^r), \  m_q' (<q^s) ,$  ($ r$ and $s$ digit numbers in base $p$ and $q$  respectively):
\[m_p'= \sum_{i=0}^{r-1}\alpha_i''p^i;
 \qquad m_q'        =\sum_{j=0}^{s-1}\beta_j''q^j\]
\item Compute by Chinese Remainder Theorem  the unique number $m'< N$ satisfying 
\[ m'\equiv m'_p\pmod{p^r};\qquad m'\equiv m'_q\pmod{q^s}\]
        For this purpose one can, using Extended Euclidean Algorithm,
        pre-compute (once)  and store $e_p, e_q$ satisfying 
\[e_p\equiv1\pmod {p^r},\ e_p\equiv 0\pmod{q^s};\qquad    e_q\equiv 1\pmod{q^s},\ e_q\equiv 0\pmod{p^r} \] 
 Then compute $m'$ by \[m' = (m'_pe_p + m'_qe_q)\pmod N\]. 
 \end{itemize}
The $m'$ obtained in the last step is the encrypted value $E_K(m)$.
\section{Comparison with AES-128}
Computations were done with SAGE Version 6.10.
The block size of AES is:
\[2^{128} = 340282366920938463463374607431768211456 \]
We searched for a number that had every prime dividing at least to the 5th
power  and \textit{close to} $2^{128}$. By trial and error we  arrived at
  \[p_1 = 163;\quad 
p_2 = 509;\quad 
p_3 = 603;\qquad 
N=p_1^5p_2^5p_3^5\]
\[N    = 340274423051874795558305386758572502851 \]
(This has the same number of decimal digits as $2^{128}$  with the most
significant four digits coinciding with it).

To  get  a  lower bound for the  number of keys 
one can compute  the number of polynomials $P_1,P_2,P_3,P_4$ of degree
at most 5, with $P_i$ involving $i$ variables, and the number of choices 
for 5-tuples of non-zero scalars. By Lemma 2, for degrees less than the finite
field order, the automorphisms would be distinct, and the table in Lemma 3
says the lower bound is

\[(603\times 509\times 163)^{209}\times (602\times
508\times162)^5.\] This works out to be of the 
order of $2^{5478}$ or $10^{1649}$.
This is much larger than the keyspace of size $2^{128}$ from AES-128,
despites a smaller block size!

\end{document}